\documentclass[final]{svjour3}
\usepackage[dvipdfmx]{graphicx}
\usepackage{rotating}
\usepackage{amssymb}
\usepackage{mathptmx}
\usepackage[numbers,sort,square]{natbib}
\usepackage{subcaption}
\usepackage{floatrow}
\usepackage{hyperref}
\usepackage{siunitx}
\usepackage{sidecap}
\usepackage[bottom]{footmisc}
\makeatletter
\sidecaptionvpos{figure}{t}
\journalname{Journal of Low Temperature Physics}

\makeatletter
\newcommand\footnoteref[1]{\protected@xdef\@thefnmark{\ref{#1}}\@footnotemark}
\makeatother

\begin{document}
\newcommand{\hdblarrow}{H\makebox[0.9ex][l]{$\downdownarrows$}-}
\title{Online Demodulation and Trigger for Flux-ramp Modulated SQUID Signals}

\author{N.~Karcher \and T.~Muscheid  \and T.~Wolber \and D.~Richter \and C.~Enss \and S.~Kempf \and O.~Sander}

\institute{N. Karcher \and T.~Muscheid  \and T.~Wolber \and O. Sander\at Institute for Data Processing and Electronics, Karlsruhe Institute of Technology, Karlsruhe, Germany,  \email{karcher@kit.edu} \and D. Richter \and C. Enss \at Kirchhoff-Institute for Physics, Heidelberg University, Heidelberg, Germany \and S. Kempf \at Institute of Micro- and Nanoelectronic Systems, Karlsruhe Institute of Technology, Karlsruhe, Germany,}


\maketitle

\begin{abstract}
Due to the periodic characteristics of SQUIDs, a suitable linearization technique is required for SQUID-based readout. Flux-ramp modulation is a common linearization technique and is typically applied for the readout of a microwave-SQUID-multiplexer as well as since recently also for dc-SQUIDs.
Flux-ramp modulation requires another stage in the signal processing chain to demodulate the SQUID output signal before further processing. For cryogenic microcalorimenters, these events are given by fast exponentially rising and slowly exponentially decaying pulses which shall be detected by a trigger engine and recorded by a storage logic. Since the data rate can be decreased significantly by demodulation and event detection, it is desirable to do both steps on the deployed fast FPGA logic during measurement before passing the data to a general-purpose processor.

In this contribution, we show the implementation of efficient multi-channel flux-ramp demodulation computed at run-time on a SoC-FPGA. Furthermore, a concept and implementation for an online trigger and buffer mechanism with its theoretical trigger loss rates depending on buffer size is presented. Both FPGA modules can be operated with up to 500 MHz clock frequency and can efficiently process 32 channels. Correct functionality and data reduction capability of the modules are demonstrated in measurements utilizing magnetic microcalorimeter irradiated with an Iron-55 source for event generation and read out by a microwave SQUID multiplexer. 

\end{abstract}

\newpage

\section{Introduction}
Cryogenic microcalorimeters based on paramagnetic or superconducting temperature sensors achieve excellent energy resolutions at low temperatures and enable groundbreaking experiments in various fields of science \cite{kempf2018b,ullom2015}.

\begin{figure}
    \centering
    \subfloat[][]{%
        \includegraphics[width=0.47\textwidth]{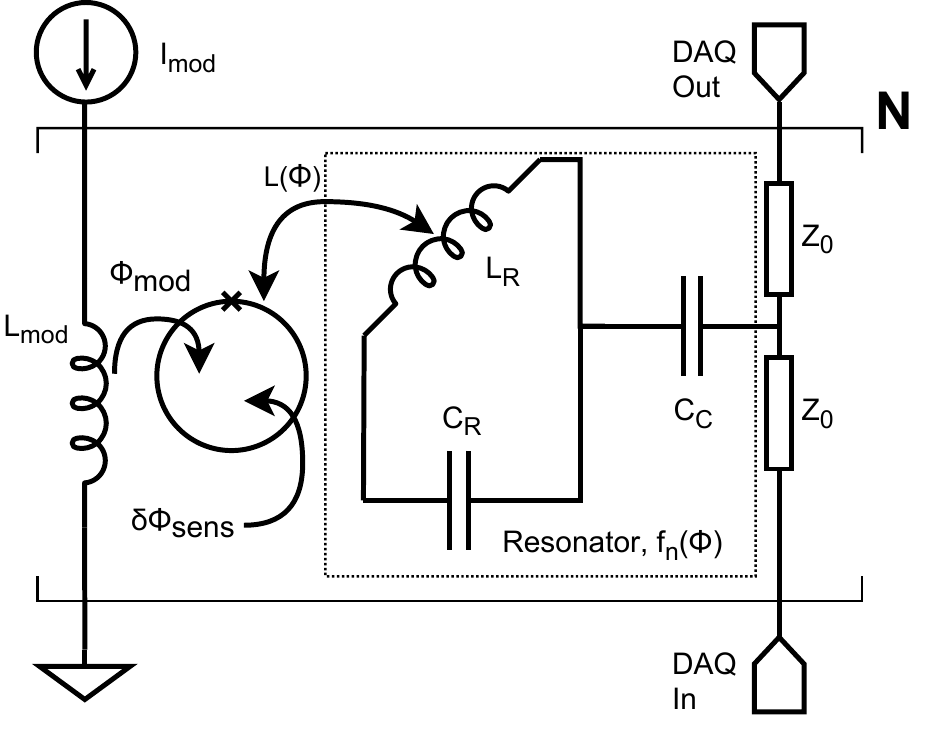}
        \label{fig:umux}%
        }%
    \subfloat[width=0.48\textwidth][]{%
        \hspace{0.04\textwidth}
        \includegraphics[width=0.39\textwidth]{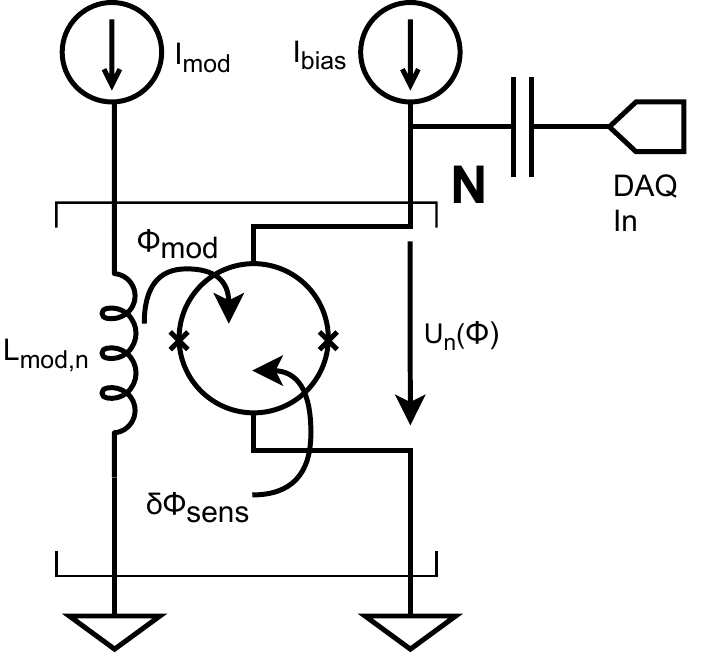}
        \hspace{0.04\textwidth}
        \label{fig:fmux}%
        }%
    \caption{Flux-ramp modulation for FDM readout concepts employing (a) rf- and (b) dc-SQUIDs. Microwave SQUID multiplexer based readout (a): A magnetic flux shifts the resonance frequency of a microwave resonator, by changing the inductance $L(\Phi)$ of an coupled rf-SQUID. A flux modulation ($I_\mathrm{mod}$,$\Phi_\mathrm{mod}$) is applied on the SQUID through $L_\mathrm{mod}$ and forms a periodic change of the resonance frequency. An additional flux from the sensor ($\delta\Phi_\mathrm{sens}$) results in a phase shift of this waveform \cite{Mates2008}. Dc-SQUID based readout (b): A magnetic flux change results in a voltage change across current biased dc-SQUID. A flux-ramp is applied through $L_\mathrm{n,mod}$ on the SQUID  ($I_\mathrm{mod}$,$\Phi_\mathrm{mod}$) and results in a periodic voltage change over the SQUID. An additional flux $\delta\Phi_\mathrm{sens}$ adds a phase to this periodic shape \cite{Richter2021}.}
    \label{fig:mux}
\end{figure}

However, system complexity with single-channel readout techniques merely scales linearly with the number of channels and increases the parasitic thermal load on the experimental platform at millikelvin temperatures for large arrays.
For this reason, frequency-division multiplexed systems based on rf-SQUIDS \cite{Mates2008,Kempf2017b} or dc-SQUID \cite{Richter2021} are used. Since the SQUID transfer functions for rf-/dc-SQUIDs are periodic, sine-like and nonlinearly dependent on the magnetic flux, the so-called flux-ramp modulation can be used for linearization with both methods\cite{Mates2012}-- in the latter, to enable multiplexing at the same time. Via an additional modulation coil, a sawtooth-shaped flux-ramp signal (periode $\tau_\mathrm{ST}$) with an amplitude of several flux quanta is induced in the SQUID. An additional flux from the sensor ($\tau_\mathrm{sig}\gg \tau_\mathrm{ST}$) acts as a quasi-static flux offset within the time frame of one flux-ramp similar to a time offset of the flux-ramp and therefore results in a phase offset of the output signal. The sensor signal can be recovered from the phase offset via demodulation. Figure \ref{fig:mux} shows how the flux-ramp modulation is combined with the multiplexing methods.

In the multiplexed readout, there is a large discrepancy between the data rate arising at the input of the AD converters (order $\mathrm{GB s^{-1}}$) and the total data rate of the finally acquired signals ($\mathrm{MB s^{-1}}$) \cite{Sander2019}. Two essential steps for reducing data rates are the demodulation of the flux-ramp, where undersampling occurs; and a triggering on events so that the idle trace can be discarded. Corresponding firmware modules has been implemented for our application, the Electron Capture in Holmium-163\cite{Gastaldo2017} experiment, they are presented in the following.


\section{Flux-ramp Demodulation}
The FPGA firmware for microwave SQUID-multiplexed signals initially requires down conversion and amplitude demodulation for channel separation \cite{Gard2018,Karcher2020}. After the filter stages a decimated, complex-valued envelope remains. By calculating the absolute value of the signal, the real-valued amplitude response can be obtained. From this point on, the processing of both multiplexing methods is similar, as the real-valued dc-SQUID signal for the flux-ramp-based multiplexing method is directly sampled by the AD converter. A major difference is that the flux-ramp-based multiplexing method a channel contains modulated signals of multiple SQUIDs, with a larger bandwidth, whereas the channel of microwave-multiplexed sensors contains a single modulated signal. 
If the frequency of the periodic oscillation is known ($f_r$), the signal can be approximately trimmed to a natural number of periods ($o_\mathrm{beg,end}$). By means of sine and cosine transformation it is mapped by a correlation to the corresponding Fourier series coefficients. Eventually the phase $\varphi_m$ for each ramp $m$ can be obtained using the arc-tangent \cite{Mates2012}: 
\begin{eqnarray}
    \varphi_m &= \arctan\biggl(\frac{\sum_{n=mM+o_\mathrm{beg}}^{m(M+N)-2-o_\mathrm{end}} s(n) \cdot \cos(2\pi\frac{f_r}{f_s} n)}{\sum_{n=mM+o_\mathrm{beg}}^{m(M+N)-2-o_\mathrm{end}} s(n) \cdot \sin(2\pi\frac{f_r}{f_s} n)}\biggl),
\end{eqnarray}
where $N$ is the length of the ramp in samples and $f_s$ the sample rate. This implies a data reduction down to the flux-ramp frequency, which is around \SI{125}{\kilo\hertz} in our case.

\begin{figure}
    \centering
    \includegraphics[width=0.90\textwidth]{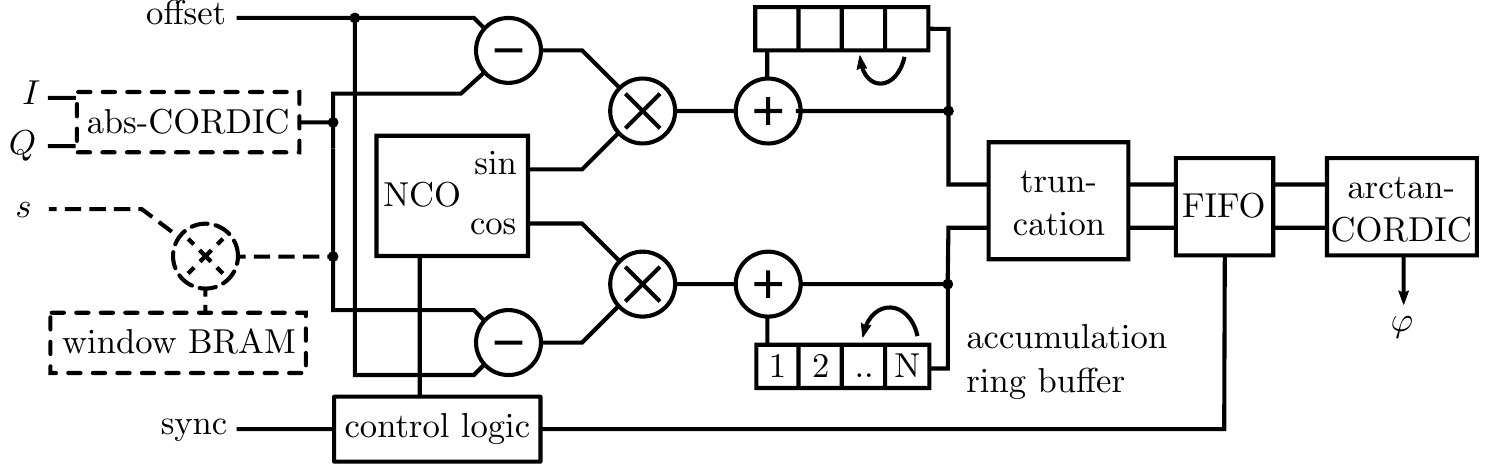}
    \caption{FPGA Firmware Block Diagram of the fluxramp demodulation module.}
    \label{fig:flux-ramp-demodulation}
\end{figure}

For resource efficiency, the implementation of the flux-ramp demodulation calculates in an interleaved, time division multiplex (TDM) fashion as shown in Fig. \ref{fig:flux-ramp-demodulation}. For the microwave SQUID-Multiplex setup, a clock frequency of \SI{500}{\mega\hertz} is used to process 32 channels at a sampling frequency of \SI{15.625}{\mega\hertz}. At the beginning, the absolute value of the input signals is formed by a pipelined\footnote{\label{pipeseq}\emph{pipelined}: one processing cycle per sample, \emph{sequential}: multiple cycles per sample} CORDIC IP core from Xilinx\textregistered{}. 
The sine and cosine values for the correlation are generated using a multi-channel numerical controllable oscillator (NCO) with direct-digital synthesis (\SI{16}{B} address and amplitude width). The computation of the correlation is performed within two DSP elements (DSP48E2). Here, the \emph{pre-adder} is used to remove a remaining DC component of the signal. Then the difference is multiplied by the sine or cosine value and added to the internal accumulator. The accumulator and offset values are stored in a ring buffer that shifts for each channel. Start and end of the accumulation is controlled by a state machine. When the correlation is complete, the accumulator values leave the ring buffer and are scaled. The scaling unit takes both accumulator values and determines from these the most significant bit of the correlation results and truncates both values accordingly. Afterwards, the values are temporarily stored in a FIFO buffer and forwarded to a sequential\footnoteref{pipeseq}  CORDIC IP core, which calculates the quotient and arc-tangent, resulting in the phase data of the channel (compare Fig. \ref{fig:measurement-full-chain}). Since the correlation period must be aligned to the flux-ramp, the ramp generator passes a synchronization pulse to the demodulator. This resets the NCO and state machine for accumulation. The flux-ramp demodulation for 32 channels with a abs-CORDIC, clocked with \SI{500}{\mega\hertz} requires 4 DSPs, 5243 LUTs and 8 BRAM units on a Xilinx\textregistered{} Zynq Ultrascale+ device\footnote{DSP: Digital signal processors; LUT: Lookup tables; BRAM: Block random access memory}.

\begin{SCfigure}[][h]
    \includegraphics[width=0.48\textwidth]{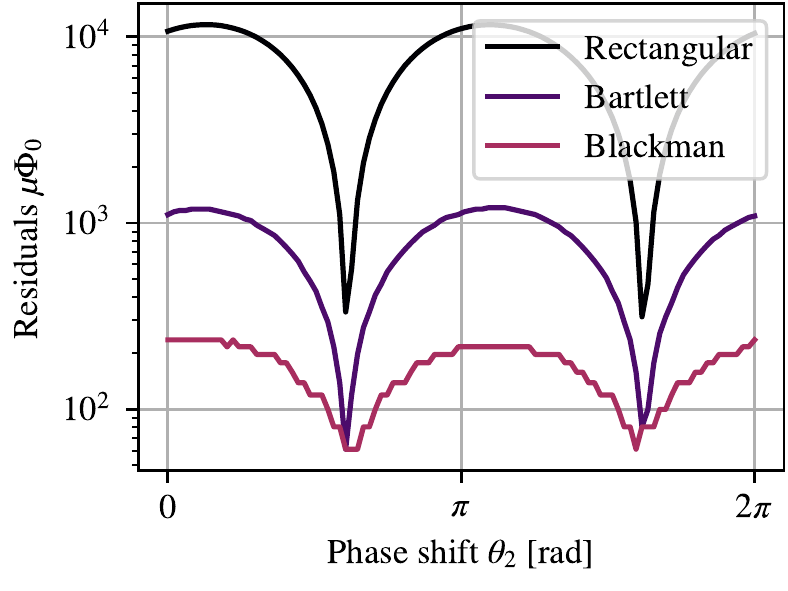}
    \caption{Simulation of the demodulation module with two sine functions as flux-ramp with 40 and 44.4 periodes (1000 samples). The figure shows the demodulation error of the first ramp for three window functions if the second signal is phase-shifted over $\mathrm{2\pi}$. Already with the low-cost barlett window isolation can be improved, in this case one order of magnitude.}
    \label{fig:flux-ramp-windowing}
\end{SCfigure}

The increased bandwidth for dc-SQUID-flux-ramp multiplexing method \cite{Richter2021} demands a high\-er signal processing sampling rate. After a decimation stage four channels are processed within the module with \SI{125}{\mega\hertz} sampling rate. The individual coupling factors of the SQUIDs lead to different modulation frequencies per channel, which makes the definition of a common correlation period difficult or even impossible. If the period can only be adjusted for one channel, spectral leakage of other channels occurs. This can be mitigated by applying a window function over the correlation period (see Fig. \ref{fig:flux-ramp-windowing}). Utilizing the windowing mechanism requires one additional DSP and flux-ramp period dependend amount of BRAM units. The total amount of resources for a four-channel module with a maximum ramp length of 1024 samples is: 3 DSP, 2218 LUTs and 9 BRAM.

\section{Event Detection}
The signal processing chain before event detection processes the channels in a TDM. After an event has been detected, an event of a channel is extracted from the TDM data stream and temporarily stored in an assigned memory slot. Eventually, the data packet is transferred by a DMA into a larger DDR memory. For efficiency reasons, it is desirable to keep the BRAM memory as small as possible. We assume a constant decay rate, with poisson-distributed events. Ideally, to capture all events, each channel is equipped with one memory slot and an event must be instantaneously fetched from the back-end. If less memory slots are provided, a loss of data might happen. A buffer overflow occurs in situations with simultaneous events on more channels than slots provided. While the decay rate is known the buffer size can be optimized such that only a reasonable amount of events is discarded. For the probability $P_b$ that an event is discarded, the Erlang-B formula from queuing theory can be used. $P_b$ for an event rate $E=\lambda \tau_c$ ($\lambda$: events per second, $\tau_c$: length of an event) and a number of limited resources or memory locations N is defined as:
\begin{equation}
    P_b = B(E,N) = \frac{E^N}{N!}\biggl/\sum_{i=0}^N \frac{E^i}{i!}.
\end{equation}
For our event rate of \SI{20}{\becquerel} of a length of \SI{3.5}{\milli\second} on active 20 channels only 5 slots must be instantiated in order to capture almost \SI{99}{\percent} of the events, comparable to sensors quantum efficiency. This is \SI{75}{\percent} less RAM than a full population. Although this model neglects time for data forwarding, Monte-Carlo simulations suggest that the effect is not significant.
\begin{figure}[b]
    \centering
    \includegraphics[width=0.90\textwidth]{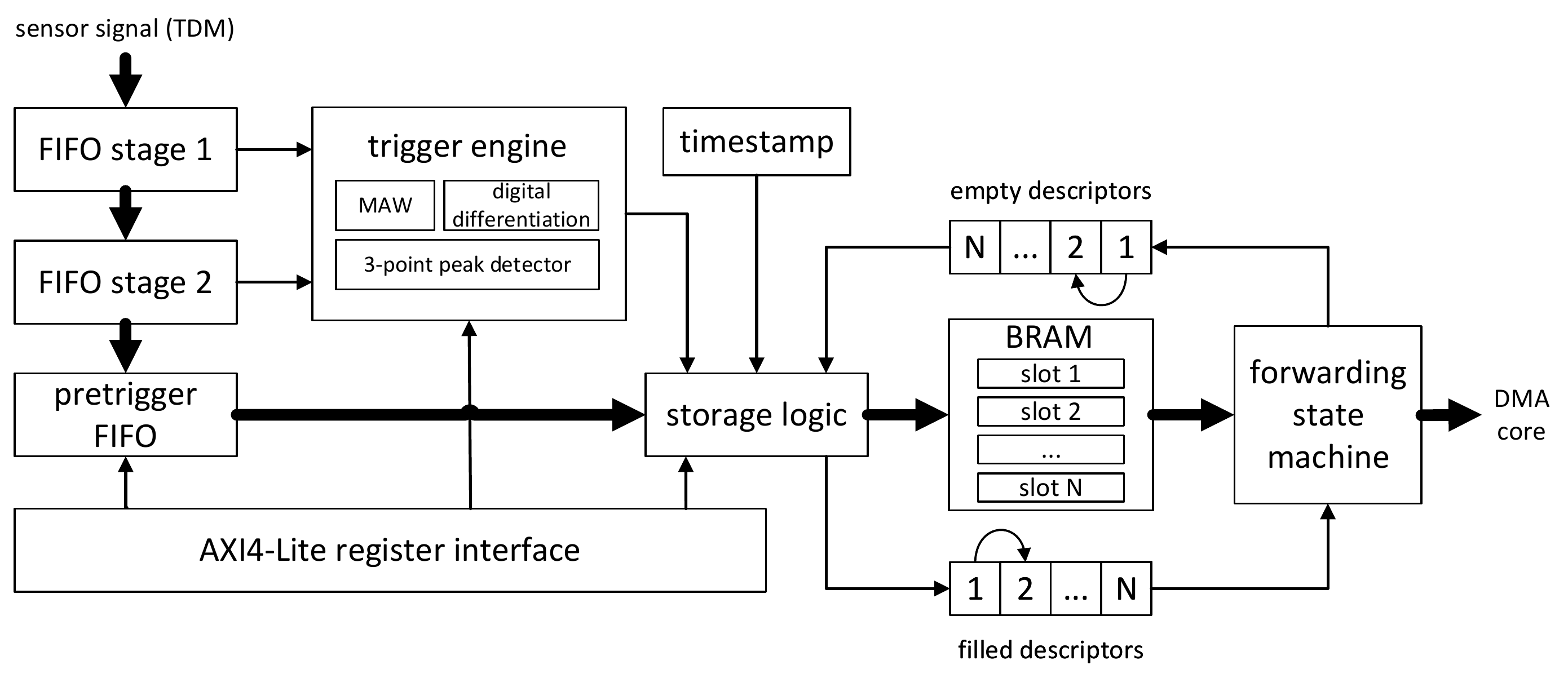}
    \caption{FPGA firmware block diagram of the event detection module.}
    \label{fig:event-detection}
\end{figure}
An overview of the event detection with its functional units is displayed in Fig. \ref{fig:event-detection}. The TDM sensor data stream first passes the trigger filter. It is implemented with two recursive moving average window (MAW) filters, each containing a shift register, a subtractor and an accumulator. The output of the filters is combined by another subtractor, that calculates the trigger input signal for the following 3-point trigger \cite{Bystricky2004}.  The trigger fires if the absolute value reaches the highest point, and it is above a predefined threshold value. Samples before the trigger time are buffered by a pre-trigger buffer, which is implemented by a synchronous FIFO buffer with variable length.

\begin{SCfigure}
    \includegraphics[width=0.483\textwidth]{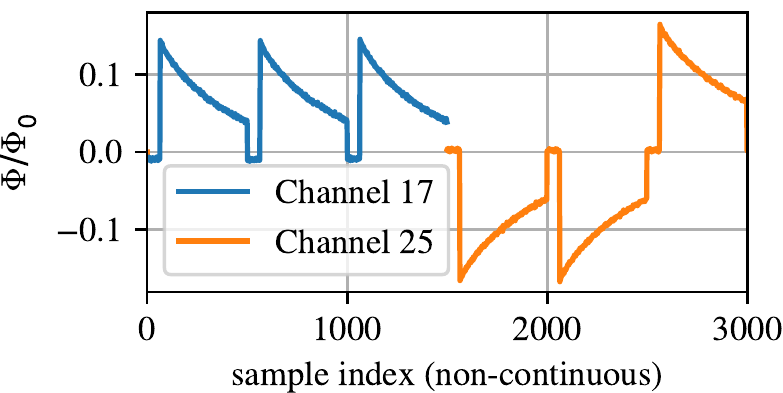}
    \caption{Photon absorption events of a $^{55}$Fe-source. Data stream of two \textmu Mux channels after frequency demultiplex, flux-ramp demodulation and event detection module. The x-axis represents scattered time intervals around the trigger times, with a sample time of \SI{8}{\micro\second}.}
    \label{fig:measurement-full-chain}
\end{SCfigure}

The event data is stored in a descriptor-based buffer, that also hands the data from the signal processing clock domain to the DMA logic clock, if required. 
The trigger state machine has a ring buffer with meta data for the current input channel in the TDM. As soon as the trigger condition is met, a timestamp is stored in the channel data and a descriptor is fetched from the free descriptor FIFO buffer. The memory area defined in the descriptor is filled with the event data for a given event length. If the trigger is fired again during saving the data, the event is marked as \emph{Pile-Up}. In the end the descriptor is pushed to the filled descriptor FIFO buffer. 
The buffer is implemented by an asynchronous two-port BRAM for the data and the two descriptor FIFO with shift registers including a clock domain crossing with handshaking. The descriptors consists of the memory address, memory length, and event meta data, such as the timestamp, trigger value and pile-up-marking.
On the DMA clock domain side, the data evacuation is controlled by a state machine. This checks each clock cycle for a new descriptor shift register. If present, the machine first passes the metadata to the data stream following the event data. After the transfer is complete, the descriptor is marked as empty and is returned into the shift register for free descriptors. The resulting data stream is the sparse phase data with a header as prefix (compare Fig. \ref{fig:measurement-full-chain}). The data reduction depends on the event rate and length. For the given parameters the reduction lies in the range of \SI{93}{\percent}. The event detection module with five slots (N=5, rounded up to N=8), a four samples MAW, a pre-trigger FIFO of 256 samples and 32 TDM channels occupy 3 DSP, 1764 LUT and 14 BRAM units.


\section{Summary}
We developed an online flux-ramp demodulation and event detection, with which individual events can be extracted from a continuous data stream of flux-ramp modulated signals.
The modules evaluate the acquired sensor data at the time of measurement, decimating the sensor signal down to the flux-ramp frequency and further reducing the data by a event rate depended factor through triggering. This corresponds to a data reduction in the order of $10^{3}$ for our application. By estimating the blocking probability through the Erlang-B formula, the amount of BRAM needed in the trigger can be greatly reduced, by \SI{75}{\percent} in our case. 
We furthermore proposed a method to suppress spectral leakage in dc-SQUID-flux-ramp multiplexed channels using window functions. The method could also improve noise characteristics and spectral leakage in \textmu Mux-systems with flux-ramp modulation. 
\begin{acknowledgements}
This work was performed in the framework of the DFG research unit FOR2202 (funding under grant No. EN29917-2). Nick Karcher acknowledges the support by the Doctoral School \emph{Karlsruhe School of Elementary and Astroparticle Physics: Science and Technology}
\\\\
The datasets generated during and/or analysed during the current study are available from the corresponding author on reasonable request.
\end{acknowledgements}

\bibliographystyle{unsrt}
\bibliography{bibliography}

\end{document}